\documentclass{cernrep}
\usepackage[colorlinks=true, linkcolor=black, citecolor=black, filecolor=black, urlcolor=blue]{hyperref}
\usepackage{fancyhdr}
\fancyhfoffset{4 mm}
\fancypagestyle{ARTTITLE}{%
\fancyhf{} 
\lhead{\footnotesize{Proceedings of the 2019 CERN--Accelerator--School course on
\it{ High Gradient Wakefield Accelerators}, Sesimbra, (Portugal)}}
\lfoot{Available online at \url{https://cas.web.cern.ch/previous-schools}}
\rfoot{\thepage\hspace*{3mm}}

   }
\begin{document}
\title{Acceleration of Electrons in Plasma}
\author{A. G. R. Thomas}
\institute{University of Michigan, Ann Arbor, Michigan, U.S.A.}

\begin{abstract}
This is brief review of acceleration of electrons in plasma wakefields driven by either intense laser pulses or particle beams following lectures at the 2019 CERN Accelerator School on plasma accelerators, held at Sesimbra, Portugal. The commonalities between drivers and their strength parameters and operating parameter regimes for current experiments in laser wakefield acceleration (LWFA) and beam driven plasma wakefield acceleration (PWFA) are summarized. Energy limitations are introduced, including the dephasing and depletion lengths for lasers, and the transformer ratio for beam driven plasmas. The concept of the wake Hamiltonian is introduced and the resulting particle orbits are identified in phase space, which illustrates how the peak energy and energy spread of accelerated electrons are determined. 
\end{abstract}

\keywords{CERN report; contribution; plasma; wakefield; acceleration.}

\maketitle
\thispagestyle{ARTTITLE}
\section{Summary of wakefield drivers}
If the limiting factor on the scale of a conventional accelerator is  breakdown of the accelerating structure, then an attractive alternative is to use a plasma. A plasma can support arbitrarily high electric fields, limited only by the obtainable charge density (and eventually, quantum effects). Longitudinal electric fields moving at the speed of light are supported in the form of relativistic electron plasma, or Langmuir waves. Generating these relativistic plasma waves requires a particle beam travelling at the speed of light propagating through the plasma. This can be any species of particle that can cause a displacement of the electrons, for example neutrinos through weak interaction or positrons through the electromagnetic force \cite{blue:pos}. For practical purposes (i.e. availability), the choice of particle beam driver  is limited to common charged particle beams (electrons or protons) \cite{Chen_PRL_1985} or photons (i.e. a pulsed laser) \cite{Tajima_PRL_1979}. In all cases, these driver beams displace background plasma electrons as they propagate to generate large amplitude plasma waves with relativistic phase velocity, as described in other lectures in this series.

\begin{table}[h]
\begin{center}
\caption{Summary of parameters in recent PWFA and LWFA experiments for comparison.}
\begin{tabular}{|c|c|c|}
\hline\hline
\textbf{Typical parameters} & \textbf{ PWFA experiments} \cite{Blumenfeld_Nature_2007,awake} &  \textbf{LWFA experiments} \cite{Albert_PPCF_2016}\\
\hline
Plasma density & $10^{16}$ cm$^{-3}$ & $10^{18}-10^{19}$ cm$^{-3}$\\
Plasma length & $\sim$m&$\sim$mm -- cm\\
Drive beam energy &$10^{10}-10^{11}$ particles @ $10$-$1000$~GeV $\sim10-10^4$ J&$1$-$10$ J laser energy \\
Drive beam duration &  10s fs (FACET), 100s ps (AWAKE)&10s fs\\
Drive beam focal size & 10s $\mu$m &  10s $\mu$m \\
\hline\hline
\end{tabular}
\end{center}
\label{table1}
\end{table}%

The main schemes of electron acceleration  are divided into "PWFA" schemes and "LWFA" schemes. "PWFA" stands for Plasma WakeField Accelerator, but generally refers to plasma wakefields generated by particle beams specifically. "LWFA" stands for Laser WakeField Accelerator. These schemes also have other acronyms in the literature, including "LPA" and "PWA". The typical parameters, including plasma density and typical spatial/temporal scales of recent PWFA and LWFA experiments are summarized in Table \ref{table1}, with PWFA being broken down into electron beam and proton beam driven cases.

\subsection{Charged particle beam and laser pulse strength parameters}
Here we review the characteristics of beam and laser pulse drivers and note some commonalities between them, especially with respect to wakefield generation. For a particle beam driver, the force that generates the perturbation in the plasma is that due to its space charge force. Assuming cylindrical symmetry, the~force on a single test (plasma) electron can be expressed in terms of the gradient of a pseudo-potential $\Psi_b \simeq \gamma_b^2\left(\phi_{b} - A_{bz} v_{b}\right)$, where $\phi_{b}$ and $A_{bz}$ are the electrostatic  and longitudinal component of the vector potentials due to the charged bunch traveling in the $+z$ direction, with $v_b$  the drive bunch velocity, $\gamma_b = \sqrt{1-v_b^2/c^2}$. The transverse force on an electron with velocity $\beta_zc$ in the $z$ (propagation) direction is therefore
$$
\mathbf{F}_{b_\bot} = -e\left(1-\beta_z\right)\nabla\Psi_b = -e \left(1-\beta_z\right)\gamma_b^2\nabla\left(\phi_{b} - A_{bz} v_{b}\right)\;.
$$

For a narrowly focused beam, outside of the beam radius $\sigma_b$ (assuming a beam profile with compact support) where the potential falls of as $\sim \ln r$, the pseudopotential, $\Psi_b \simeq \Psi_0\ln(r/\sigma_b)$, is related to the integrated charge per unit length by  
$$
\Psi_0 \simeq \frac{1}{\varepsilon_0}\int_0^{\sigma_b} \rho_b(r,z) r dr\;,
$$
where $\rho_b(r,z)$ is the drive beam particle charge. The longitudinal force is $1/\gamma_b^2$ weaker in strength, $F_{b\parallel} = -e\nabla\Psi_b / \gamma_b^2$, and so the transverse expulsion of electrons dominates. The constant $\Psi_0$ can be expressed in dimensionless form using the usual relativistic plasma normalization; $v\rightarrow v/c$, $x\rightarrow x\omega_p/c$, $t\rightarrow \omega_p t$, $p\rightarrow p/mc$, $E\rightarrow eE/m_ec\omega_p$, $\rho\rightarrow \rho/\rho_0$ etc., where $\omega_p=\sqrt{e\rho_0/m_e\varepsilon_0}$ is the  plasma frequency. Using these quantities, $\Lambda_0 = e\Psi_0/m_ec^2 = \int_0^{\sigma_b\omega_pc} \frac{\rho_b(r,z)}{\rho_0}  (\omega_pr/c) d(\omega_pr/c)$.

The strength parameter for particle beam drivers is therefore $\Lambda_0$, with $\Lambda_0\ll1$ meaning the wake is in the linear regime and for $\Lambda_0\gg1$ it is in the highly nonlinear ``blowout'' regime. The blowout radius of the plasma wake in the strongly nonlinear regime with a particle beam driver, where the space charge repulsion balances the attractive force of the ion channel, can be shown to be $k_pr_b\approx 2\sqrt{\Lambda_0}$ \cite{Lu_POP_2006}. 

For a laser pulse, the force that generates the perturbation in the plasma is that due to its ponderomotive force,
$$
\mathbf{F}_{p} = -\frac{e^2}{2\langle\gamma\rangle m_e}\nabla \langle A^2\rangle\;,
$$
where the angle brackets indicate the cycle average over the fast timescale oscillations in the laser pulse \cite{StartsevEA_PRE_1997}. Again, expressed in the dimensionless units above, the relevant  strength parameter is $a_0 = e|A|/m_ec$,  with $a_0\ll1$ and $a_0\gg1$ indicating the linear and highly nonlinear ``bubble'' regimes. The~radius of the plasma ``bubble'' in the strongly nonlinear regime with a laser pulse driver, where the ponderomotive force balances the attractive force of the ion channel, can be shown to be $k_pr_b\approx 2\sqrt{a_0}$ \cite{Lu_POP_2006}. These results are summarized in Table \ref{table2}.
\begin{table}[htp]
\caption{Summary of strength parameters for PWFA and LWFA }
\begin{center}
\begin{tabular}{|c|c|c|}
\hline\hline
\bfseries Parameter & \bfseries LWFA & \bfseries PWFA\\
\hline
Strength parameter & $a_0$ & $\Lambda_0$ \\
Linear regime & $a_0\ll1$& $\Lambda_0\ll1$\\
Nonlinear regime &$a_0\gg1$& $\Lambda_0\gg1$\\
Wake potential amplitude (linear regime) & $\propto a_0^2$&$\propto \Lambda_0$\\
Wake potential amplitude (nonlinear regime) &$\propto a_0$&$\propto \Lambda_0$\\
Nonlinear regime radius & $k_pr_b\approx 2\sqrt{a_0}$&$k_pr_b\approx 2\sqrt{\Lambda_0}$\\
\hline\hline
\end{tabular}
\end{center}
\label{table2}
\end{table}%

\begin{figure}[htbp]
\begin{center}
\includegraphics[width=0.5\textwidth]{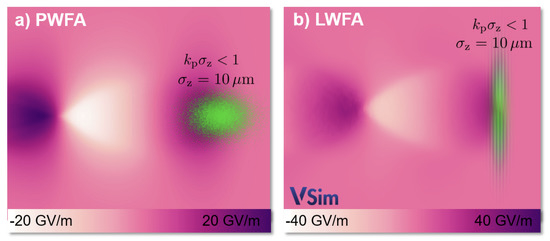}
\caption{PIC-simulation visualization of electron-driven PWFA (a) and laser-driven LWFA (b) in the blowout/bubble regime, respectively. The driver beam is shown in green and propagates to the right, expels plasma electrons (not shown) and thus generates strong trailing electric decelerating/accelerating fields. Figure from \emph{Fundamentals and Applications of Hybrid LWFA-PWFA}, B. Hidding et al., Appl. Sci. 2019, 9(13), 2626.}
\label{hiddingfigure}
\end{center}
\end{figure}

For a short driver (relative to the plasma period), behind the driver the plasma will respond the~same regardless of the driver. Hence, although beam and laser pulse drivers generate very different plasma perturbations in the vicinity of the driver, behind the driver where the gradients are accelerating, the plasma perturbation will be generally quite similar regardless of the driver type, as illustrated in Fig.~\ref{hiddingfigure}. Hence we can discuss acceleration by a plasma wakefield in quite general terms.
\pagebreak
\subsection{Accelerating and focusing wake phases}
\begin{figure}[htbp]
\begin{center}
\includegraphics[width = 0.4\textwidth]{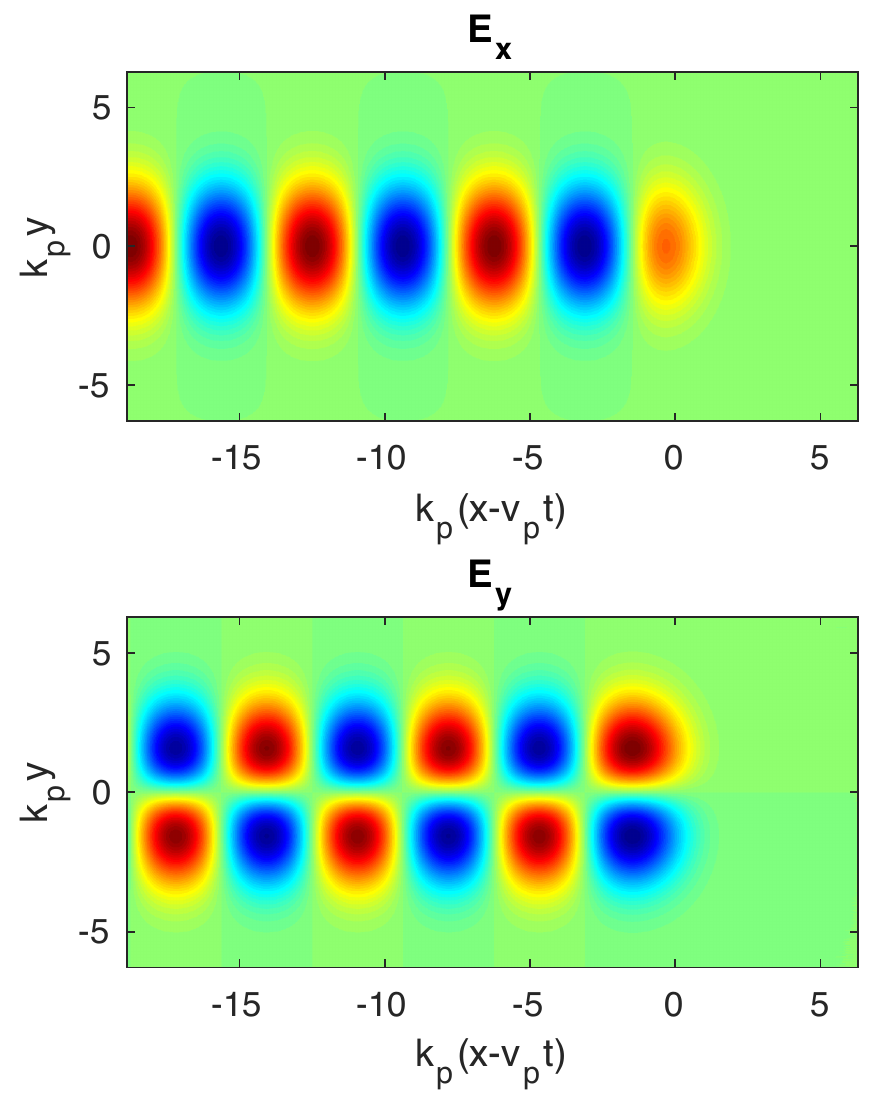}
\caption{Electric fields of a wake generated by a laser driver with $a_0 = 0.1$.}
\label{wake1}
\end{center}
\end{figure}

A wakefield has oscillating fields that demonstrate both accelerating / decelerating regions and focusing / defocusing regions. In the linear regime, there are four $\pi/2$ phases with each combination of accelerating / decelerating and focusing / defocusing equally distributed. For acceleration of a beam, it is necessary to be in the phase of the wake that is both focusing and accelerating. Figure \ref{wake1} shows the longitudinal ($x$) and transverse ($y$) electric fields of a wake generated by a laser driver with $a_0 = 0.1$, illustrating this four-fold symmetry.

\begin{figure}[htbp]
\begin{center}
\includegraphics[width = 0.4\textwidth]{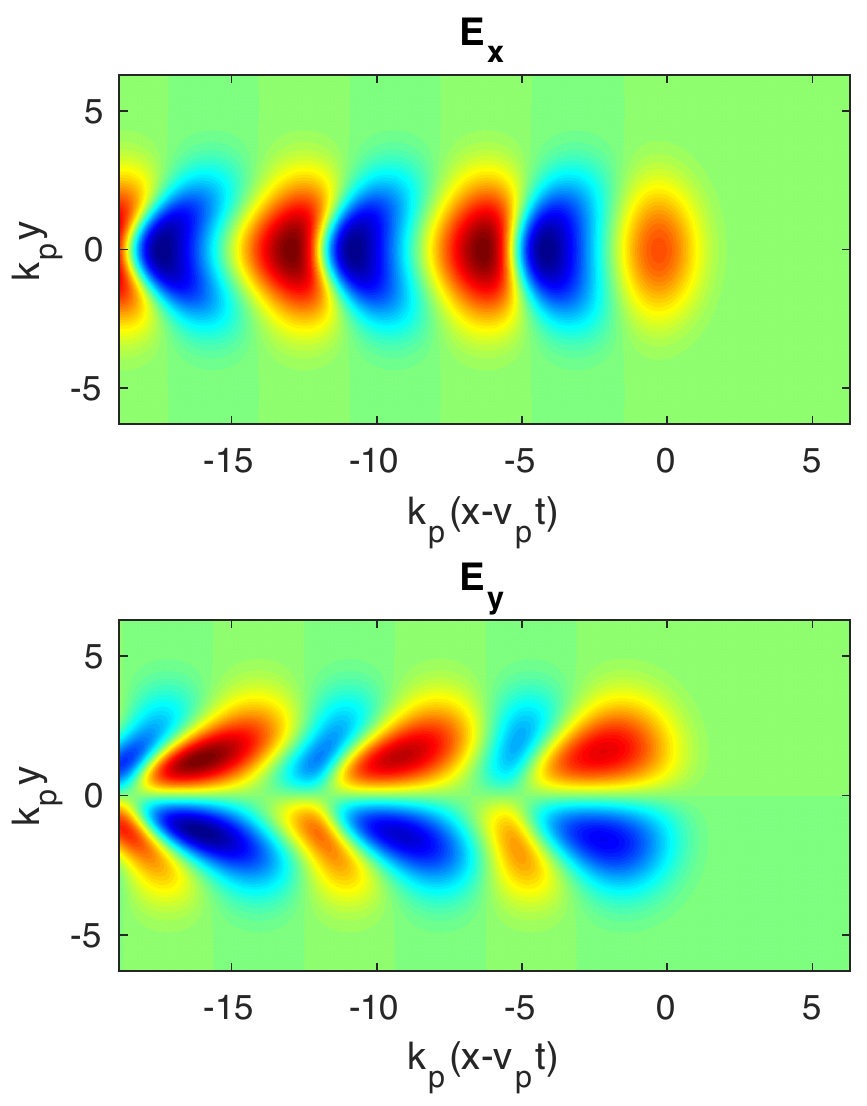}
\caption{Electric fields of a wake generated by a laser driver with $a_0 = 1$.}
\label{wake2}
\end{center}
\end{figure}

In the nonlinear regime, as illustrated in Fig. \ref{wake2} where $a_0=1$, the wake accelerating / decelerating regions and focusing / defocusing regions become very asymmetric, with the fields being primarily focusing for electrons. This is obviously very advantageous for electron acceleration, but is a significant disadvantage for positron acceleration where the focusing phase becomes very small. Other more complex schemes have been proposed to overcome this. See other lectures for details.
\section{Limits on Maximum energy gain of an electron beam}
All plasma wakefield accelerators involve the generation of a plasma wave with relativistic phase velocity by a perturbing object (laser pulse, charged particle beam) traveling at near light speed. The generated plasma wakefield may accelerate an electron beam  known as a ``witness beam''. 
These lectures will concentrate on the considerations of wakefields generated by a general relativistic perturbing object (laser pulse, particle beam etc.) with \emph{an approximately constant velocity and which doesn't change in amplitude}. In reality, the driver evolves as it propagates, due to nonlinear refractive index or beam head erosion etc. effects, which are left for later lectures. 

The phase velocity of the wakefield structure will be dictated by the velocity of the relativistic driver, which for a charged particle beam driver of energy $\gamma_bm_ec^2$ will be $v_b = \sqrt{1-1/\gamma_b^2}$ and for a~laser pulse driver with group velocity (envelope velocity) $v_g = \sqrt{1 - \omega_p^2/\omega_0^2}$ for linear dispersion, where $\omega_0$ is the laser (angular) frequency, we can assign an effective Lorentz factor $\gamma_g = \omega_0/\omega_p$. The~laser evolution is in reality quite complicated, and simulation studies have implied an effective Lorentz factor for laser propagation of $\gamma_g = \omega_0/\sqrt{3}\omega_p$ \cite{Decker_PRL_1994} due to pulse front erosion. The details of this are left for other lectures, but it is enough for us to say that we can define an effective Lorentz factor for the plasma wake structure generated by the relativistic driver, whether a charged particle beam or laser pulse, which we denote $\gamma_p$.

\begin{figure}[htbp]
\begin{center}
\includegraphics[width=0.9\textwidth]{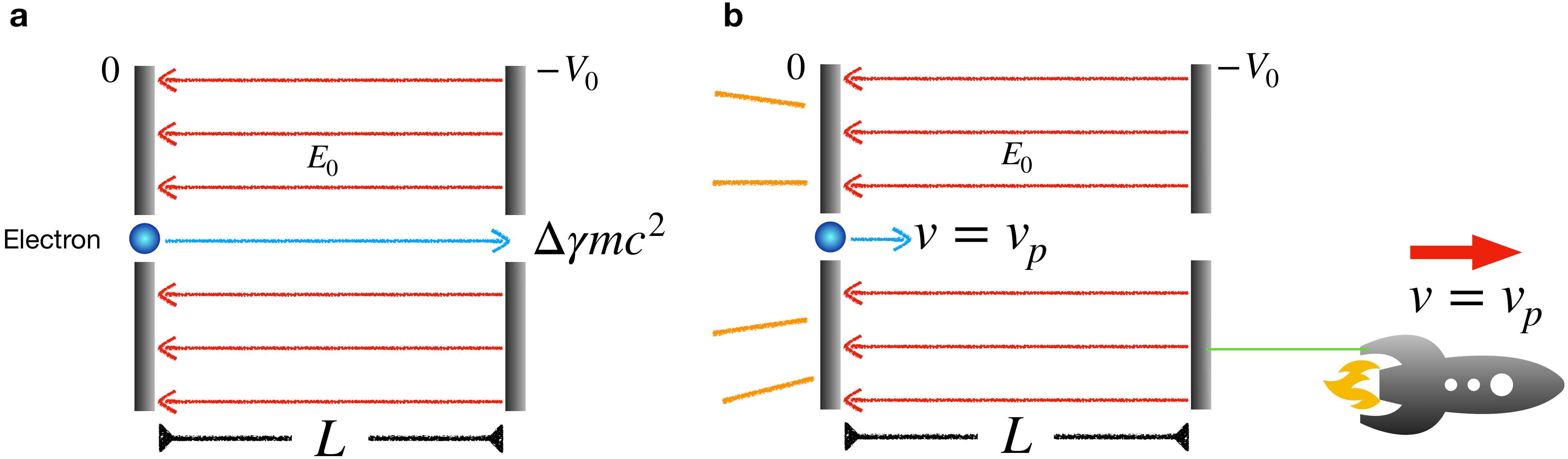}
\caption{Cartoon showing the effect of a moving accelerating structure on the maximum energy gain. (a) A pair of parallel plates. (b) Parallel plates being towed by a rocket.}
\label{cartoon}
\end{center}
\end{figure}

 As a cartoon example of how energy gain is affected by acceleration in a moving wave structure, consider an electron gaining energy between a pair of parallel plates with a potential difference of $-V_0$, as shown in Fig. \ref{cartoon}(a). The energy gain of the electron crossing the gap is, trivially, $\Delta\gamma mc^2 = eV_0$  Now imagine we attach the parallel plates to a rocket that is pulling them at a constant speed $v = v_p$. What is the maximum energy gain we can achieve? Now as the electron passes through the plates, they move along with it, thus lengthening the time the electron spends in the accelerating fields and therefore increasing the energy gain.

There are two things that can occur to limit the energy gain in this cartoon example. Either the~electron reaches the other side of the parallel plates or the rocket runs out of fuel. These have analogues for acceleration by a plasma wakefield. In the former case,  the limit is known as \emph{dephasing}, since the electron has exited the phase of the accelerating moving structure. In the latter case, the limit is known as \emph{depletion}. In a plasma accelerator, the \emph{dephasing} limit is typically associated with laser drivers, since the~group velocity of the laser driver has a relatively big difference from the speed of light in vacuum and so the accelerated particle eventually outruns the pulse. The \emph{depletion} limit is typically associated with particle beam drivers, as experiments have been run with ultra-relativistic drivers such that the witness electron beam does not catch up with the driver before the driver is significantly depleted of energy. Laser drivers can also be limited by depletion of the laser pulse energy under certain conditions. 
 
\subsection{Dephasing of the electron beam}
In the case that the wake has a relatively slow phase velocity\footnote{This means that the difference between the wake velocity and the speed of light, $\Delta v/c = 1-v_p/c$ is not small compared with $c/\omega_pL$, where $L$ is the plasma length. If $\Delta v/c\ll c/\omega_pL$, then the phase slippage relative to the speed of light frame is negligible since it is small compared to the plasma-wave length-scale.}, the accelerated electrons cannot indefinitely stay in phase with the accelerating phase of the plasma wave due to the small, but not insignificant, difference in velocity between the pulse and electron bunch $v_z-v_p\neq 0$.  This means that the electron bunch will eventually out-run the accelerating part of the wakefield, which places an upper limit on the~maximum energy gain by these electrons in a single (uniform density) plasma stage. The length over which this occurs is known as the \emph{dephasing length} $L_d$.

The simple model we will use to develop scalings for this maximum energy is to consider the~electron beam to have been ``injected'' somehow, which is to say that it is initialized with a velocity equal to the phase velocity of the plasma wake, $v_{p}$ at some point behind the driver. We do not consider the details of how it reached these conditions, just that it initially has $v_z = v_p$ somewhere in the plasma wake. A useful tool is to use the coordinates of the ``wake frame''; this is a frame of reference moving at the phase velocity of the relativistic plasma wave. This frame is not a Lorentz boosted frame but simply a (Galilean) coordinate transform from  $(x,y,z,t)$ to new coordinates $(x,y,\xi,\tau)$ where  $\xi = z - v_pt$ is the \emph{wake phase}  and $\tau=t$. Using the chain rule for partial derivatives:
\begin{equation*}
\frac{\partial}{\partial t}=\frac{\partial\tau}{\partial t}\frac{\partial}{\partial \tau}+\frac{\partial\xi}{\partial t}\frac{\partial}{\partial \xi}
\end{equation*}
\begin{equation}
\;\;\;=\frac{\partial}{\partial \tau}-v_{ph}\frac{\partial}{\partial \xi}
\end{equation}
\begin{equation*}
\frac{\partial}{\partial z}=\frac{\partial\tau}{\partial z}\frac{\partial}{\partial \tau}+\frac{\partial\xi}{\partial z}\frac{\partial}{\partial \xi}
\end{equation*}
\begin{equation}
=\frac{\partial}{\partial \xi}\;\;\;\;\;\;\;
\end{equation}

If the driver is slowly evolving, the time derivative $\partial/\partial\tau\rightarrow 0$, and so all quantities depend on $\xi$ only. For a general wakefield, with longitudinal electric field $E_z(\mathbf{r},t)$ that is used to accelerate the electron beam, if the driver is non-evolving and cylindrically symmetric then the the field can be expressed as $E_z(r,\xi)$. Provided the electrons stay close to the axis (in the linear regime; this restriction is not necessary in the blowout regime, because the accelerating field is uniform in the radial direction), the accelerating field can be expressed as a function of a single parameter only,  $E_z = E_z(\xi)$. 


The equation for longitudinal energy gain $\gamma_\parallel mc^2$ of an electron in this field, using 
$$
\frac{d\xi}{dt} = v_z - v_p\;, 
$$
where $v_z$ is the longitudinal velocity of the accelerating electron, is
$$
\frac{d}{d\xi} \gamma_\parallel m_ec^2 = \frac{-eE_z(\xi) }{1- v_p/v_z}\;.
$$
If we assume that the electron is traveling at very close to the speed of light, $v_z\approx c$, this can be expressed as 
$$
\frac{d}{d\xi} \gamma_\parallel m_ec^2 \simeq -e\gamma_p^2(1+\beta_p)E_z(\xi) \;,
$$
which has the solution for the energy gained by an electron in the wakefield of
$$
\Delta \gamma_\parallel m_ec^2 \simeq -\gamma_p^2(1+\beta_p) \int_{\xi_i}^{\xi_f} eE_z(\xi) d\xi\;,
$$
where $\xi_i$ is the initial phase of the electron and $\xi_f$ is the final phase.  

The assumption that we may replace $v_z$ with $c$ in these calculations is valid for the following reasons: Since for a longitudinally directed electron, $v_z/c = \sqrt{1-1/\gamma^2}$ and the plasma wave phase velocity is $v_p/c = \sqrt{1-1/\gamma_p^2}$,  $1-v_p/v_z\simeq (1/2\gamma_p^2)(1-\gamma_p^2/\gamma^2)$ to leading order in $1/\gamma_p^2$ and $1/\gamma^2$.	\newline A 50 MeV electron has $\gamma^2\approx10^4$, whereas $\gamma_p^2\sim100$ for typical experimental densities of \newline $10^{18}-10^{19}$ cm$^{-3}$. A trapped electron must have $\gamma > \gamma_p$ in any case.  Hence, this approximation is reasonable for most of the acceleration process. There are ways of violating this. For example, for large amplitude oscillations about the axis, the longitudinal velocity component will be reduced on average because $v_z = p_z/\gamma m$ and $\gamma$ contains the transverse momentum due to the oscillations.
We can define the~average electric field experienced by a particle as 
$$
\overline{E}_z =  \frac{\int_{\xi_i}^{\xi_f} E_z(\xi) d\xi}{\Delta\xi}\;,
$$
where $\Delta\xi = {\xi_f}-{\xi_i}$ is the wake phase the particle travels in the acceleration process.  Hence the~longitudinal energy gain can be expressed as
$$
\Delta \gamma_\parallel m_ec^2 \simeq 2\gamma_p^2 e|\overline{E}_z|\Delta\xi \;.
$$ 
The distance in the laboratory frame over which this acceleration occurs can be calculated via $\Delta\xi = \int_{t_i}^{t_f} (v_z - v_p )dt$, where $t_i$ and $t_f$ correspond to the times where the particle is at wake phases $\xi_i$ and $\xi_f$ respectively. This can be expressed as 
$$
\Delta\xi = \int_{t_i}^{t_f} v_z\left(1 - \frac{v_p}{v_z} \right)dt\;.
$$
Using the above expansion, neglecting the $\gamma_p^2/\gamma^2$ term, 
$$
\Delta\xi \approx \frac{1}{2\gamma_p^2}\int_{t_i}^{t_f} v_z dt = \frac{1}{2\gamma_p^2}L_{deph}\;,
$$
where $L_{deph}$ is the length over which the acceleration occurs, known as the \emph{dephasing length}. We can see that by this definition, in general:
$$
\Delta \gamma_\parallel m_ec^2 \simeq e|\overline{E}_z| L_{deph} \;.
$$
Note that this expression is quite general; we have not said anything about the form of the electric field shape, but assume that the motion is highly relativistic and paraxial. We may also view this from the~point of view of the wake potential. Since 
$$
E_z(\xi) = -\frac{\partial \phi}{\partial \xi}\;,
$$
the longitudinal energy gain can be expressed as 
$$
\Delta \gamma_\parallel m_ec^2 \simeq \gamma_p^2(1+\beta_p) \int_{\xi_i}^{\xi_f} e\frac{\partial \phi}{\partial \xi} d\xi = \gamma_p^2(1+\beta_p) e[\phi(\xi_f) - \phi(\xi_i)]\;,
$$
where it is assumed that the function $g(x,y)$ of integration (of a partial differential) is zero. Hence, although the maximum energy gain will depend on the details of the wakefield shape,  the benefit of this general analysis is that it is clear that regardless of the details of the wakefield structure, the energy gain for a relativistic wake phase velocity $\beta_p\rightarrow 1$ will be approximately $\Delta \gamma_\parallel m_ec^2\approx2\gamma_p^2e\Delta \phi$ \cite{Esarey_RMP_2009}, where $\Delta\phi$ is the potential difference between the maximum potential in the wake and the potential at the phase when the electron is ``injected'' (with $v_z = v_p$). 

\begin{figure}[htbp]
\begin{center}
\includegraphics[width=\textwidth]{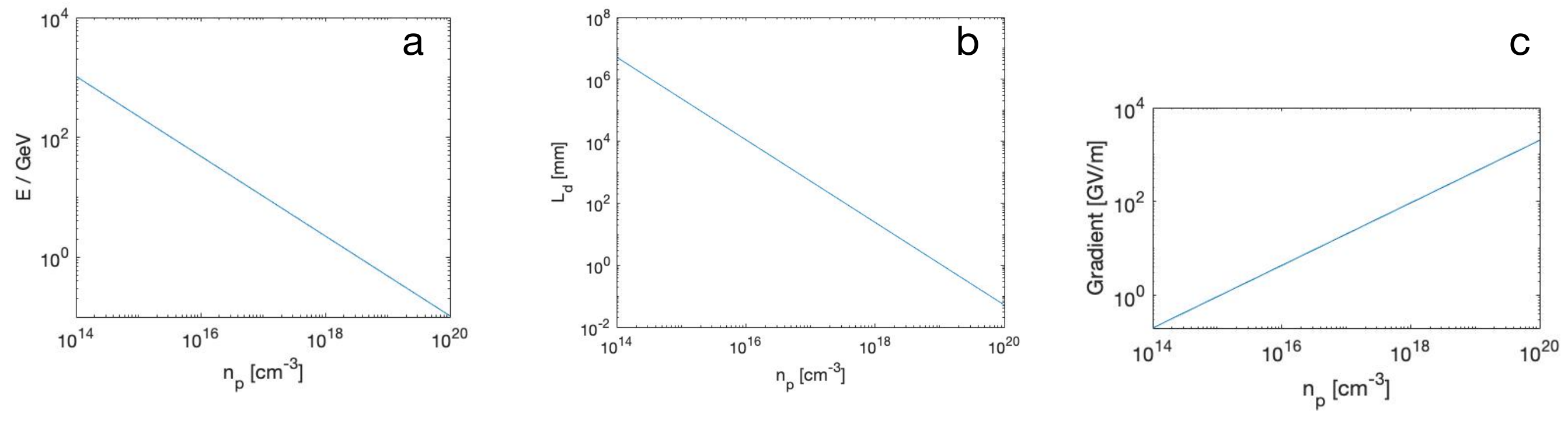}
\caption{Scalings with plasma number density for LWFA in the 3D nonlinear regime for an 800 nm laser with matched conditions. (a) The energy gain (b) dephasing (accelerator) length (c) effective accelerating gradient.}
\label{figscale}
\end{center}
\end{figure}

 In Table \ref{lutable}, the resulting scalings for the energy gain in the linear, nonlinear (1D) and nonlinear (3D) regimes are summarized. The details of the wakefield generation in linear and nonlinear 1D and 3D regimes by beam and laser drivers are covered in other lectures. These scalings in the 3D nonlinear regime have been confirmed by simulations \cite{Lu_PRZ_2007} and numerous experiments \cite{Mangles_IEEE_2008}. Figure \ref{figscale} illustrates the scaling for maximum energy gain, accelerator length (dephasing limited) and effective accelerating gradient for an 800 nm laser driver under matched conditions. Note that ``advanced accelerators'' are defined as having accelerating gradients greater than 1 GeV/m, so this sets a practical lower limit on the~density and therefore maximum energy \emph{in a single stage}. 
\begin{table}[htp]
\caption{Scalings for laser wakefield acceleration, based on table I in ref. \cite{Lu_PRZ_2007} in addition to modifications from \cite{Esarey_RMP_2009}, which are consistent with the above relations. The parameters in the columns are the normalized field strength $a_0$, normalized \emph{averaged} accelerating gradient $|\overline{E}_z|$, the effective wavelength of the wake $\lambda_W$, dephasing length $L_{deph}$, pump depletion length $L_pd$, effective wake phase velocity $\gamma_p$, and the dephasing and depletion limited electron energy gains,  $\Delta \gamma_{\parallel, deph}$ and $\Delta \gamma_{\parallel, pd}$, respectively.\label{lutable}}
\begin{center}
\begin{tabular}{|c|c|c|c|c|c|c|c|c|}
\hline\hline
\bfseries Regime & \pmb{$a_0$} & \pmb{$\frac{e|\overline{E}_z|}{mc\omega_p} $} & \pmb{$\frac{\lambda_W\omega_p}{c}$} & \pmb{$\frac{L_{deph}\omega_p}{c}$} & \pmb{$\frac{L_{pd}\omega_p}{c}$} & \pmb{$\gamma_p$} & \pmb{$\Delta \gamma_{\parallel, deph}$} & \pmb{$\Delta \gamma_{\parallel,pd}$}\\
\hline
Linear &$a_0\ll1$&$\frac{a_0^2}{2\pi}$&$2\pi $&$2\pi\frac{\omega_0^2}{\omega_p^2}$&$\frac{1}{a_0^2}\frac{\omega_0^2}{\omega_p^2}\omega_p\tau$&$\frac{\omega_0}{\omega_p}$&$a_0^2\frac{\omega_0^2}{\omega_p^2}$&$\frac{\omega_0^2}{\omega_p^2} \frac{\omega_p\tau}{2\pi}$\\
\hline
Nonlinear (1D) &$a_0\gg1$&$\frac{a_0}{2}$&$4 a_0$&$4a_0^2\frac{\omega_0^2}{\omega_p^2} $&$\frac{\omega_0^2}{\omega_p^2}\omega_p\tau$&$\sqrt{a_0}\frac{\omega_0}{\omega_p}$&$2a_0^3\frac{\omega_0^2}{\omega_p^2}$&$a_0\frac{\omega_0^2}{\omega_p^2}\omega_p\tau$\\
\hline
Nonlinear (3D) &$a_0\gg1$&$\frac{\sqrt{a_0}}{2}$&$4 \sqrt{a_0}$&$\frac{4\sqrt{a_0}}{3}\frac{\omega_0^2}{\omega_p^2}$&$a_0\frac{\omega_0^2}{\omega_p^2}\omega_p\tau$&$\frac{1}{\sqrt{3}}\frac{\omega_0}{\omega_p}$&$\frac{2}{3}a_0\frac{\omega_0^2}{\omega_p^2}$&$a_0^{3/2}\frac{\omega_0^2}{\omega_p^2}\omega_p\tau$\\

\hline\hline
\end{tabular}
\end{center}
\end{table}%

\subsection{Depletion of a laser driver}
Both a laser driver and beam driver lose energy in generating a wakefield, which also provides a limitation on acceleration length. For the laser, we may define a \emph{pump-depletion length}, $L_{pd}$, which is the length over which the driver loses energy and beyond which it may be assumed the driver is no longer able to generate a large amplitude wakefield. 

An estimate of the pump depletion length for an arbitrary driver can be made by equating the~energy in the driver with the electromagnetic energy transferred to the generated wake field. The maximum electromagnetic energy  of the wake in a region of thickness $\delta z$ near the axis where only the $E_z$ component of the fields is non-zero  over a small cross-sectional area $\delta A$ is 
$$
\delta U_W \approx \frac{1}{2}\varepsilon_0 \overline{E_z^2}\delta A \delta z\;.
$$
Hence the total energy transferred to the plasma wake over a length $L_{pd}$ is
$$
U_W \approx \frac{1}{2}\varepsilon_0 \overline{E_z^2}\delta A L_{pd}\;.
$$

We can equate this with the driver energy over the same small cross-sectional area $\delta A$, assuming a driver energy density $\eta$ and pulse length $\tau$
$$
U_d \approx \eta \delta A c\tau\;,
$$
such that
$$
L_{pd} = \frac{2\eta c\tau}{\varepsilon_0 \overline{E_z^2}}\;.
$$
If the driver is a laser pulse, the energy density is 
$$
U_{d}({\rm laser}) =  \frac{1}{2}\varepsilon_0 E_L^2 =  \frac{1}{2}\frac{\varepsilon_0 m^2c^2\omega_0^2}{e^2} a_0^2
$$
and therefore 
$$
\frac{L_{pd}\omega_p}{c} = \frac{\omega_0^2}{\omega_p^2}\frac{a_0^2}{ \overline{(eE_z/mc\omega_p)^2}}\omega_p\tau\;,
$$
which leads to the scalings given in Table \ref{lutable}.


\section{Particle orbits in a plasma wakefield}
\subsection{Wake Hamiltonian}
The textbook Hamiltonian for an electron in electromagnetic potentials $\mathbf{A}(\mathbf{x},t)$ and $\phi(\mathbf{x},t)$  is 
$$
H = \sqrt{(mc^2)^2 + (\mathbf{P}-e\mathbf{A})^2c^2} -e\phi\;,
$$
where $\mathbf{P} = \gamma m\mathbf{v}+e\mathbf{A}$ is the canonical momentum, and  the Lorentz factor of the electron in the field is $\gamma = \sqrt{1 + (\mathbf{P}-e\mathbf{A})^2/m^2c^2}$. 

For the wakefield, we would like to express the Hamiltonian in terms of the more natural wake coordinate $\xi = z-v_pt$, since if the driver is non-evolving, all quantities will vary as $x,y,\xi$ only. Through a canonical transformation using a generating function 
$$
F_2 = \mathbf{P}\cdot \mathbf{x} -  v_p\int(P_z - eA_z) dt
$$
we can re-express this in terms of the coordinates $x,y,\xi,t$ as the \emph{wake Hamiltonian}
$$
H_W = \sqrt{(mc^2)^2 + (\mathbf{P}-e\mathbf{A})^2c^2} -e\psi - v_pP_z\;,
$$
where $\psi = \phi - v_pA_z$. This expression is similar to that found in \cite{Esirkepov_PRL_2006}, but includes the longitudinal component of the vector potential, which generates the azimuthal magnetic field in 3D  wakefield structures. We can show that this Hamiltonian is consistent with the equations of motion by explicitly calculating Hamilton's equations,
$$
\dot{\mathbf{x}} = \frac{\partial H_W}{\partial \mathbf{P}} = \frac{\mathbf{P}-e\mathbf{A}}{\sqrt{(mc)^2 + (\mathbf{P}-e\mathbf{A})^2}} - v_p = \mathbf{v}-v_p\hat{z}\;,
$$ 
where $\mathbf{v} = (\mathbf{P} - e\mathbf{A})/\gamma m$, which the kinetic velocities as expected with $\dot{\xi} = v_z - v_p$, and 
$$
\dot{\mathbf{P}} = -\frac{\partial H_W}{\partial \mathbf{x}} = \frac{\partial e\mathbf{A}}{\partial\mathbf{x}}\cdot\frac{\mathbf{P}-e\mathbf{A}}{\sqrt{(mc)^2 + (\mathbf{P}-e\mathbf{A})^2}}+ \frac{\partial e\psi}{\partial\mathbf{x}} =  \frac{\partial e\mathbf{A}}{\partial\mathbf{x}}\cdot\mathbf{v} +\frac{\partial e\phi}{\partial\mathbf{x}} - v_p\frac{\partial eA_z}{\partial\mathbf{x}}\;,
$$
which is the same as that arising from the standard electromagnetic Hamiltonian result (which can be shown equivalent to $\mathbf{E}+\mathbf{v}\times\mathbf{B}$) but with an additional term $- v_p\frac{\partial eA_z}{\partial\mathbf{x}}$ which arises due to the change of coordinates from $x,y,z,t\rightarrow x,y,\xi,t$.
 If we assume that there is no explicit time dependence, which is equivalent to saying that the driver is non-evolving, then this Hamiltonian is conserved. Hence,
\begin{equation}
\frac{H_W}{mc^2} =h_0  = \gamma - \beta_pu_z - \Psi\;,\label{eqh0}
\end{equation}
is a useful constant of motion, where $ \Psi = {e\psi}/{mc^2}$ and $u_z = P_z/mc$. Note that $e\psi + P_zv_p = e\phi + p_zv_p$, so this can also be written as $\gamma(1 - \beta_p\beta_z) - \frac{e\phi}{mc^2}$, where $\beta_z = v_z/c$. This last expression predicts the~same energy gain as given earlier, since it implies that the change between initial, $i$, and final, $f$, points in the~trajectory is 
 $$
 \frac{e\Delta \phi }{mc^2}= \gamma_f(1 - \beta_{z,f}\beta_p) -  \gamma_i(1 - \beta_{z,i}\beta_p)\;,
 $$
 which, using the same $\beta_z\rightarrow 1$ approximation as previously rearranges to 
 $$
 \Delta\gamma \approx 2\gamma_p^2  \frac{e\Delta \phi}{mc^2} \;.
 $$
\subsection{Wake phase space}
The wake Hamiltonian can be used to illustrate the different trajectories in the wake \emph{phase space}. The~phase space is a map of  all possible particle coordinates, but we restrict it for practical purposes (to show as a 2D image) to being the possible coordinates of particles in the phase space $(P_z, \xi)$. For simplicity, consider a plane wave, nonevolving driver, such that $\Psi$ and $\mathbf{a} = e\mathbf{A}/mc$ only depend on $\xi$. Transverse canonical momentum is conserved in this case, so $P_x = P_y = 0$, and by choice of gauge $\nabla\cdot\mathbf{A} = 0$, $a_z = 0$. The wake Hamiltonian is (in normalized form and setting $H_w = h_0mc^2$), therefore 
$$
h_0 = \sqrt{1+ a(\xi)^2 + u_z^2} -\Psi(\xi)- \beta_pu_z\;.
$$

\begin{figure}[htbp]
\begin{center}
\includegraphics[width = 0.5\textwidth]{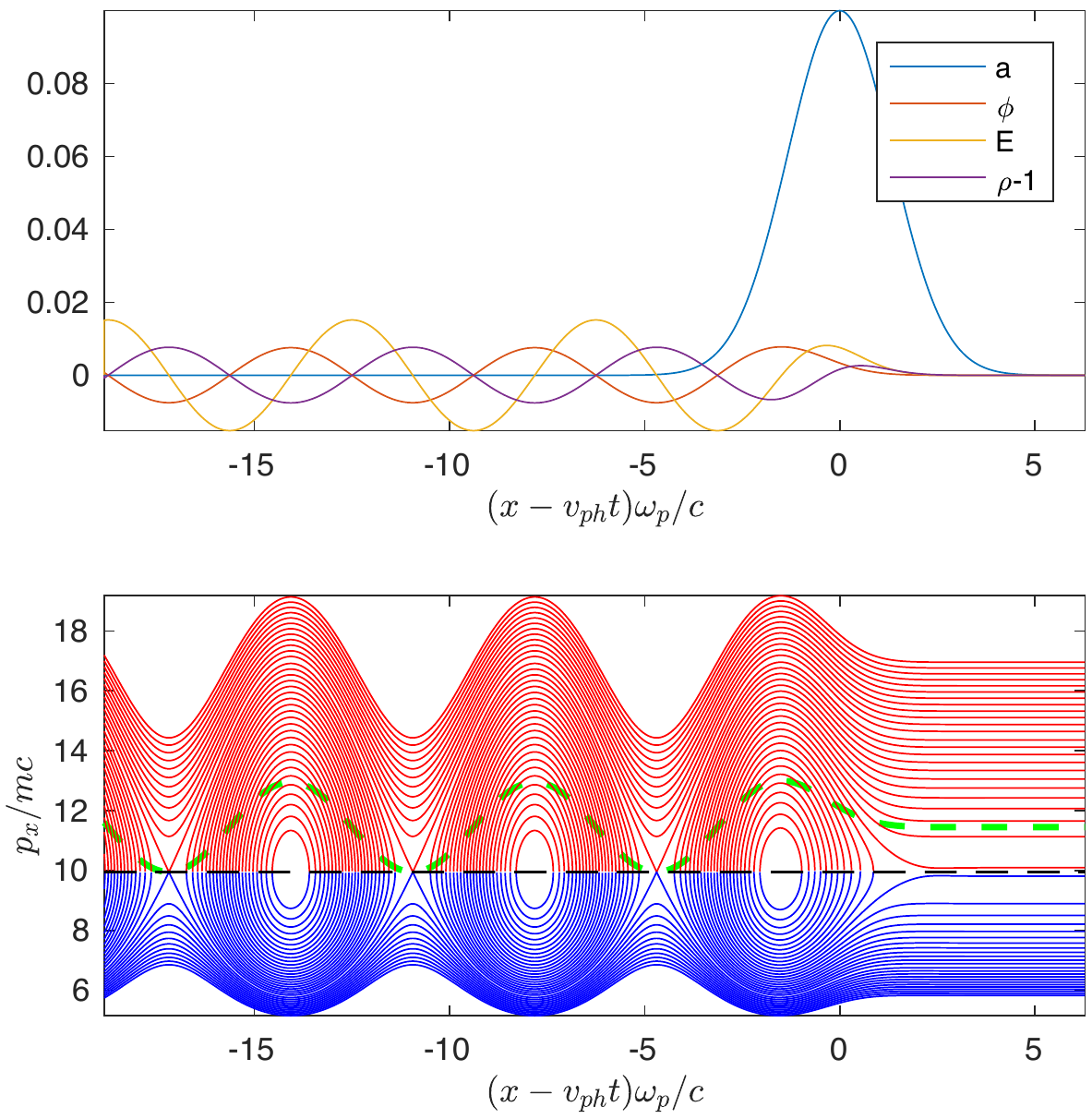}
\caption{(top) Laser field $a$, wake potential $\phi$, electric field $E$ and density perturbation $\delta\rho$ for a laser driver with amplitude $a_0 = 0.1$. (bottom) Resulting phase-space trajectories (red and blue). The green dotted line indicates the wake potential.}
\label{phase1}
\end{center}
\end{figure}

Each particle trajectory is defined by a value for $h_0$, and so by rearranging Eq. (\ref{eqh0}), we can solve for the longitudinal momentum of the particle as a function of $\xi$, as
\begin{equation}
u_z = \gamma_p^2(h_0 + \Psi(\xi))\left(\beta_p\pm\sqrt{1 - \frac{1+a(\xi)^2}{\gamma_p^2(h_0+\Psi(\xi))^2}}\right)\;. \label{eqntrak}
\end{equation}
The $+$ and $-$ components of this stitch together at $h_0 + \Psi = 1/\gamma_p + a^2$, which from the Hamiltonian can be seen to be the turning points when $u_z = \gamma_p\beta_p$. Trajectories which include $u_z = \gamma_p\beta_p$ will consist of closed orbits in phase space, whereas for those that do not touch $u_z = \gamma_p\beta_p$ there are two solutions, one passing in the forward direction and one passing in the backward direction. A trajectory that is stationary in the \emph{laboratory frame} will be appearing to propagate backwards in the  \emph{wake frame}. A particle that is travelling at the wake phase velocity in the  \emph{laboratory frame} will appear stationary in the  \emph{wake frame}.

To demonstrate the properties of these solutions, following Ref. \cite{Esirkepov_PRL_2006} we calculate numerical solutions of the nonlinear Poisson equation for plasma waves, which stated without derivation\footnote{See other lectures for derivation of nonlinear plasma wakes.} is
$$
\frac{\partial^2\Psi}{\partial\xi^2} = \gamma_p^2k_p^2\left(\beta_p\left(1 - \frac{1+a^2}{\gamma_p^2(1+\Psi)^2}\right)^{-1/2}-1\right)\;,
$$
using a gaussian envelope driver field of the form
$$
a^2 = a_0^2\exp\left(-8\ln2\frac{k_p^2\xi^2}{\pi^2}\right).
$$
Figure \ref{phase1} indicates these phase space orbits for such a calculation for a laser driver with amplitude $a_0 = 0.1$. (top) Laser field $a$, wake potential $\phi$, electric field $E$ and density perturbation $\delta\rho$. (bottom) Resulting phase-space trajectories (red and blue). The green dotted line indicates the wake potential. The~red and blue trajectories correspond to the $+$ (red) and $-$ (blue) components of the expression given in Eq.~(\ref{eqntrak}). Where they meet, you can observe continuous orbits circulating in phase space. These trajectories correspond to the orbits of electrons that are \emph{trapped} in the wake.

\begin{figure}[htbp]
\begin{center}
\includegraphics[width = 0.5\textwidth]{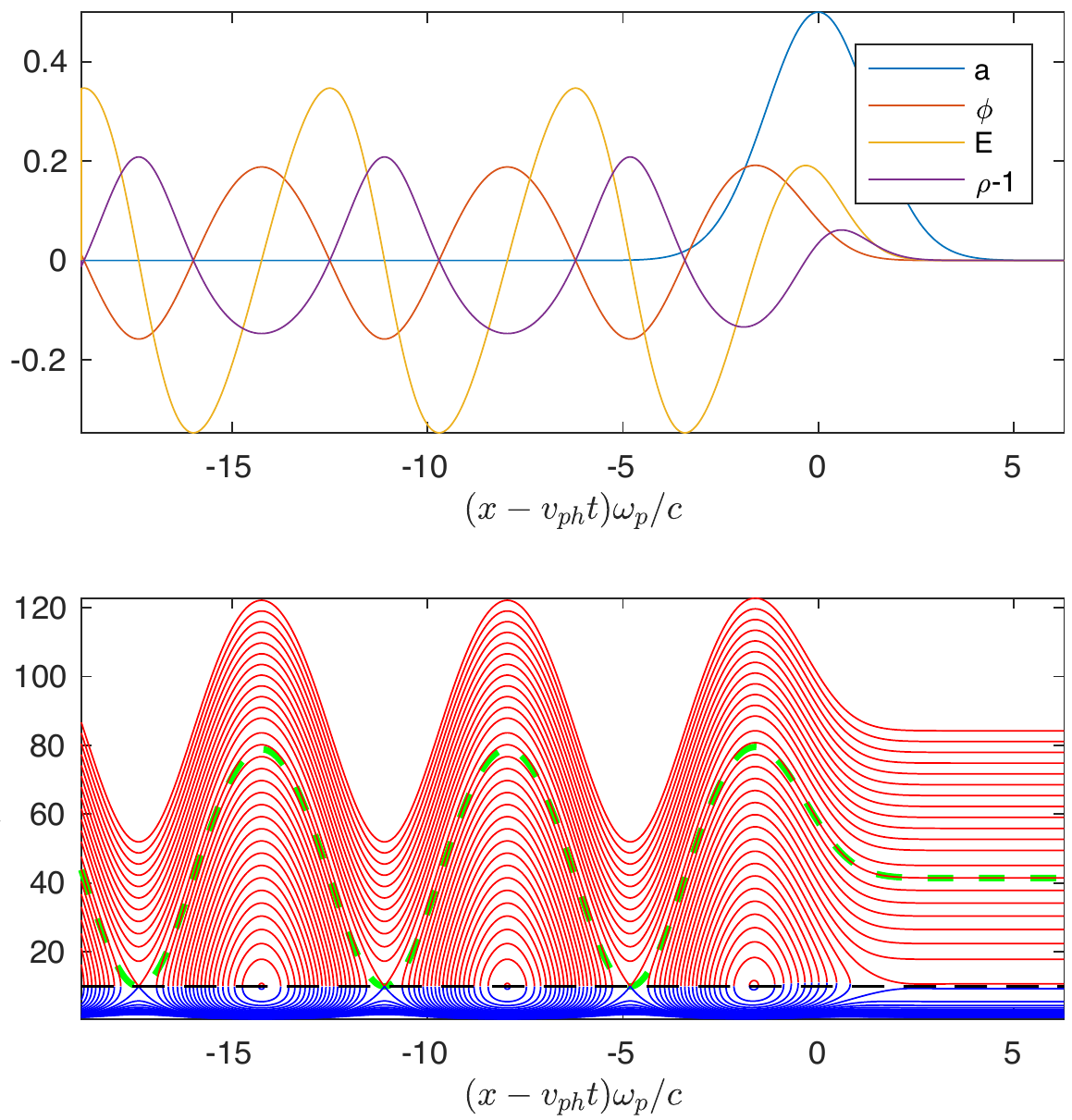}
\caption{(top) Laser field $a$, wake potential $\phi$, electric field $E$ and density perturbation $\delta\rho$ for a laser driver with amplitude $a_0 = 0.5$. (bottom) Resulting phase-space trajectories (red and blue). The green dotted line indicates the~wake potential.}
\label{phase2}
\end{center}
\end{figure}

\begin{figure}[htbp]
\begin{center}
\includegraphics[width = 0.5\textwidth]{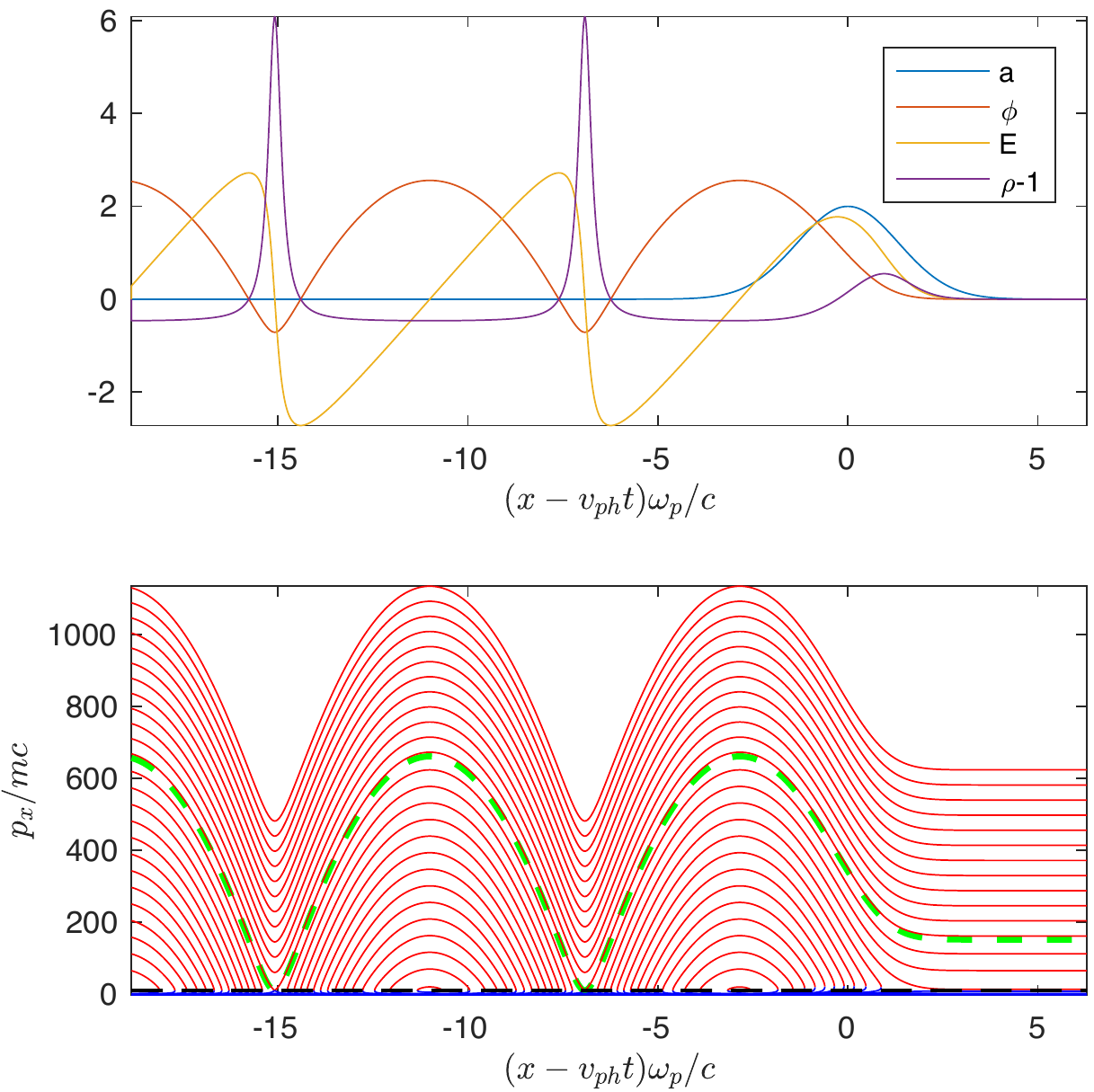}
\caption{(top) Laser field $a$, wake potential $\phi$, electric field $E$ and density perturbation $\delta\rho$ for a laser driver with amplitude $a_0 = 2$. (bottom) Resulting phase-space trajectories (red and blue). The green dotted line indicates the~wake potential.}
\label{phase3}
\end{center}
\end{figure} 

As the wake driver amplitude increases, the generated wake potential changes from being sinusoidal in the linear regime to a parabolic shape in the strongly nonlinear regime. Figures \ref{phase2} and \ref{phase3} show the phase space orbits for laser drivers with amplitude $a_0 = 0.5$ and $a_0= 2$ respectively, indicating this transition to parabolic orbits. As the amplitude increases, the range of trapped orbits increases to include those electrons initially at rest in the laboratory frame. This corresponds to the phenomenon of wavebreaking, as clearly if large numbers of background particles are trapped, the wave structure will be destroyed since the framework of this problem is that there is a  wave made up of the background electrons that is unperturbed by the presence of `witness particles' that are accelerated on trapped orbits in the phase space. 

\begin{figure}[htbp]
\begin{center}
\includegraphics[width = 0.5\textwidth]{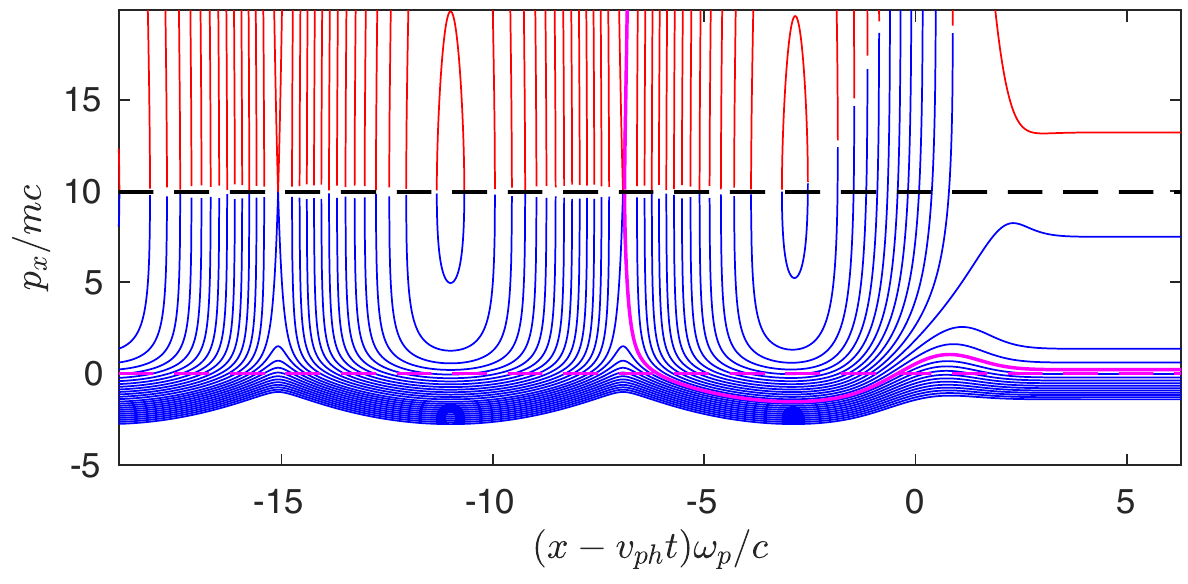}
\caption{Zoomed in region of figure \ref{phase3} showing the resulting phase-space trajectories (red and blue) near $u_z = 0$. The purple dashed line indicates $u_z = 0$. The purple solid line indicates electrons with zero momentum far ahead of the laser which end up near the separatrix.}
\label{phase4}
\end{center}
\end{figure}

Figure \ref{phase4} shows the same phase space as Fig. \ref{phase3} but zoomed in to look at the detail near $u_z = 0$. The purple dashed line indicates $u_z = 0$. The purple solid line indicates electrons with zero momentum far ahead of the laser which end up near the separatrix. This indicates that for a large enough amplitude wave, wavebreaking can occur. If there is an initial distribution of particle momenta, the particles that are on average moving forward may be trapped. Alternatively, if electrons are `born' at rest within the wake structure, at a phase corresponding to a closed orbit, it will be trapped -- this is the case for ionization injection \cite{Pak_PRL_2010,McGuffey_PRL_2010}, where inner shell electrons are bound closely to the ions until ionized within the wake structure.

\subsection{Phase rotation and extraction}
Once trapped, electrons are accelerated by following the phase space orbits to the turning point of the~potential, which corresponds to the dephasing length in the laboratory frame. This represents the maximum energy gain. For useful applications, the spectral shape should be a peaked distribution around a single energy with small spread.

The simplest way to produce electrons at a single energy is to accelerate them with identical force over an identical distance. That is to say, in the absence of \emph{beam loading effects} monoenergetic electrons with a small energy will be produced if a bunch of electrons occupy a configuration space region that is small compared to the electric field gradient, and are all trapped within a period short compared with the~overall acceleration time.

\begin{figure}
\begin{center}
\includegraphics[width=0.7\textwidth]{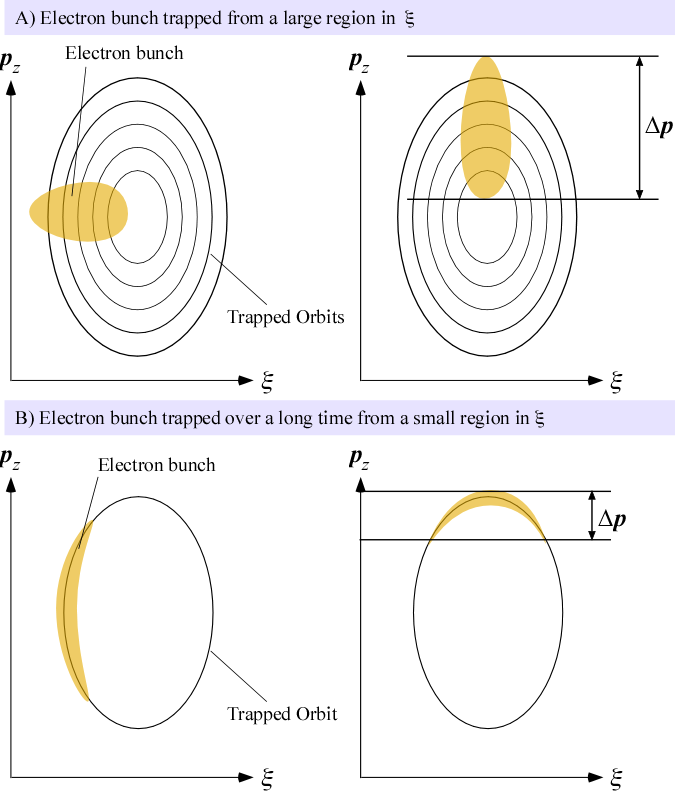}
\caption{Momentum-phase space diagrams for monoenergetic bunch production.}
\label{orbitz}
\end{center}
\end{figure}

The dephasing length has already been discussed with respect to the maximum energy gain from the wakefield accelerator, but it also affects the energy spread. In a non-evolving wake, electrons can occupy closed orbits in momentum-phase space, as illustrated in Fig. \ref{orbitz}. They are accelerated and subsequently decelerated. Injected electrons will all follow very similar orbits, and thus a bunch of electrons along an orbit will rotate in phase space, resulting in a small energy spread. The energy spread of the electron bunch can be minimized if it is extracted at the point of highest energy, the turning point of the potential well. This is the dephasing length. Extraction of the electron bunch at the dephasing length means terminating the plasma at this length.

Of course, the wakefield is not usually non-evolving, and so propagating for longer than a dephasing length is likely to result in increased energy spread compared to the original bunch. At high densities, trapping continues for long times and many orbits are taken by the electrons and so a broad energy spread is the likely result. Such a spectrum looks very similar to experimentally measured spectra. 

In Figure \ref{orbitz}, panel (A) shows that even if the trapped electron distribution has a short temporal distribution, if trapped over a large range of phases, since the electrons all follow the closed phase space orbits, they will end up occupying an undesirable large range of energies. Panel (B) shows that if the~electrons are trapped over a small range of wakes phases but for a long period of time, the energy spread will be compressed. Indeed, even if the injection is continuous, a quasi-monoenergetic electron spectrum can be achieved. 

\begin{figure}
\begin{center}
\includegraphics[width=5in]{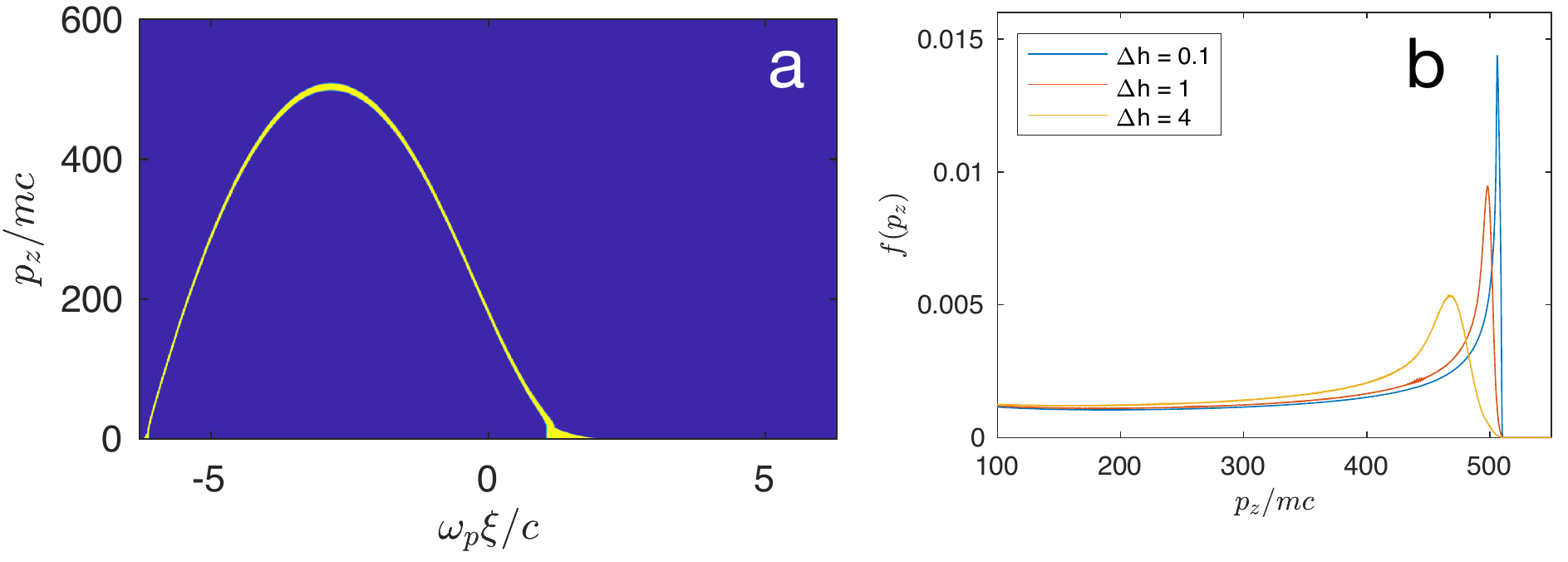}
\caption{(a) Momentum-phase space diagram showing filled orbit with width $\Delta h_0= 0.1$. (b) Effective electron spectrum calculated from filled phase-space.}
\label{momp}
\end{center}
\end{figure}

In Figure \ref{momp} (a),  the phase space density for particle orbits that are completely filled within a~range of $h_0$ values, $\Delta h_0$. This is equivalent to a situation where there is continual injection of electrons at the~rear of the first wakefield period over the small range of wake phases in a non-evolving wake. 

Each electron trajectory contributes a $\delta(P_z - P_z(\xi, h_0))$ to the overall distribution, where $\delta(x)$ is the Dirac distribution, and so for a given injection distribution (i.e., how the electrons are distributed on the phases space orbits), $f_i(h_0)$, the overall phase-space distribution will be $f(P_z,\xi;h)= \int_{-\infty}^{\infty} f(h_0)\delta(P_z - P_z(\xi, h_0))dh_0$. Therefore, the spectrum is given by:
$$
f(P_z) =  \int_{\xi_i}^{\xi_f}\int_{-\infty}^{\infty} f(h_0)\delta(P_z - P_z(\xi, h_0))d h_0 d\xi\;.
$$
In Figure \ref{momp}  (b), the blue line shows the electron spectrum that results by integrating over all phases using this expression with $f(h) = f_0 \exp(-h^2/\Delta h^2)$, which shows a characteristic quasi-monoenergetic peak that arises naturally because of this phase rotation effect. The red and yellow lines show increasing values of $\Delta h_0$ and its effect on the electron spectrum.

\pagebreak
\subsection{Transformer ratio}
\begin{figure}
\begin{center}
\includegraphics[width=0.6\textwidth]{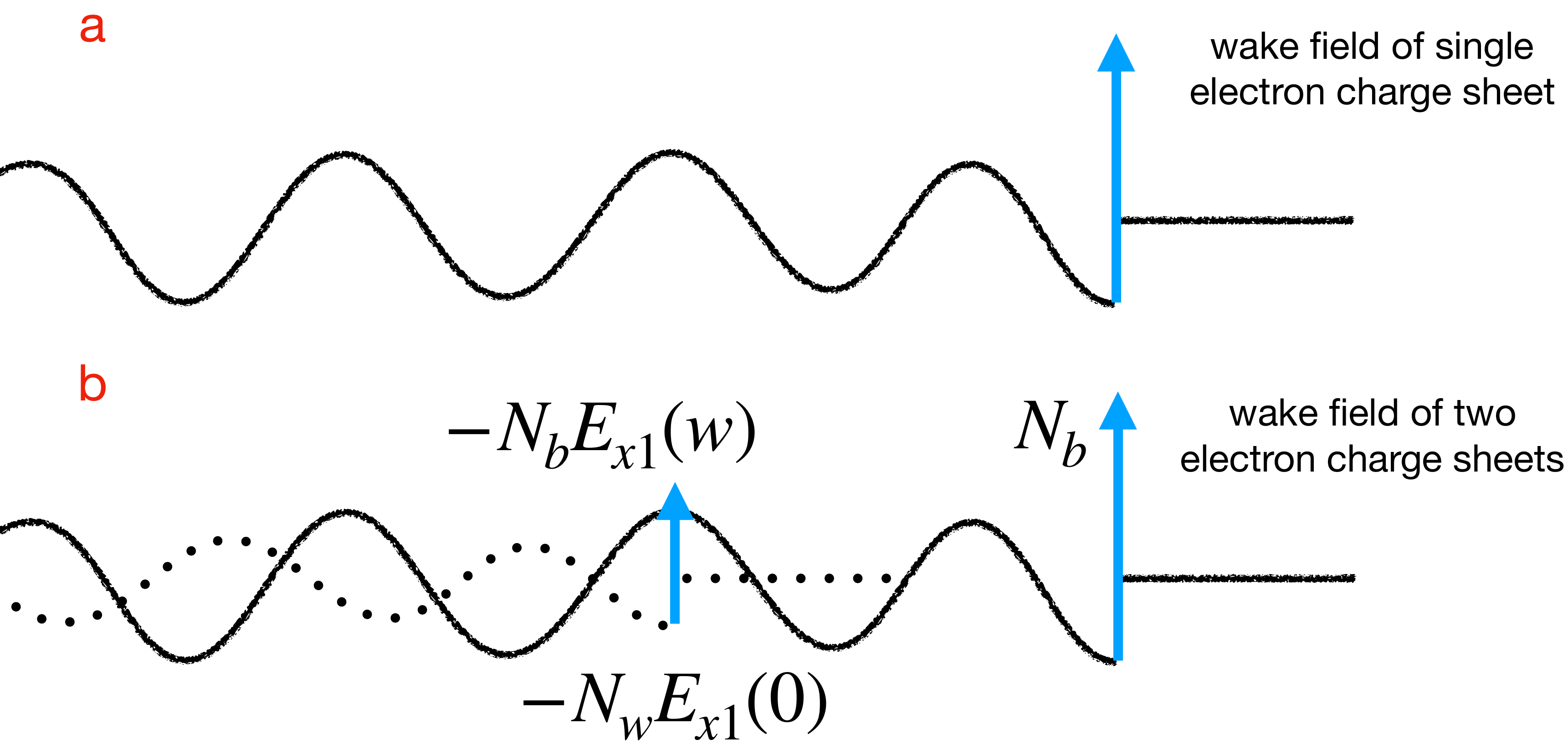}
\caption{(a) Wakefield generated by a short driver. (b) Wakefield generated by a short driver and a short witness bunch.}
\label{waketrans}
\end{center}
\end{figure}

In PWFA \cite{Chen_PRL_1985}, one important definition is the \emph{transformer ratio}, which is a limitation on the energy transfer from one beam to another. We accelerate a ``witness'' bunch in the wake of the first ``drive'' bunch, but the witness bunch  generates its own wakefield. Following the derivation in reference \cite{Ruth_PA_1985} The total wakefield  is a linear superposition of wake contributions from both the drive bunch and witness bunch. Assume that there are two short bunches located at wake phases $\xi = z-v_pt = 0$ and $\xi = w$, with number densities for the drive beam,  $N_b$, and witness beam, $N_w$. Each \emph{individual} electron induces an \emph{identical} wakefield with electric field $E_{z1}(\xi)$. The total wakefield is a linear superposition of $E_{z1}(0)$ and $E_{z1}(w)$, as shown in Fig. \ref{waketrans}. 

\begin{figure}[htbp]
\begin{center}
\includegraphics[width = 0.3\textwidth]{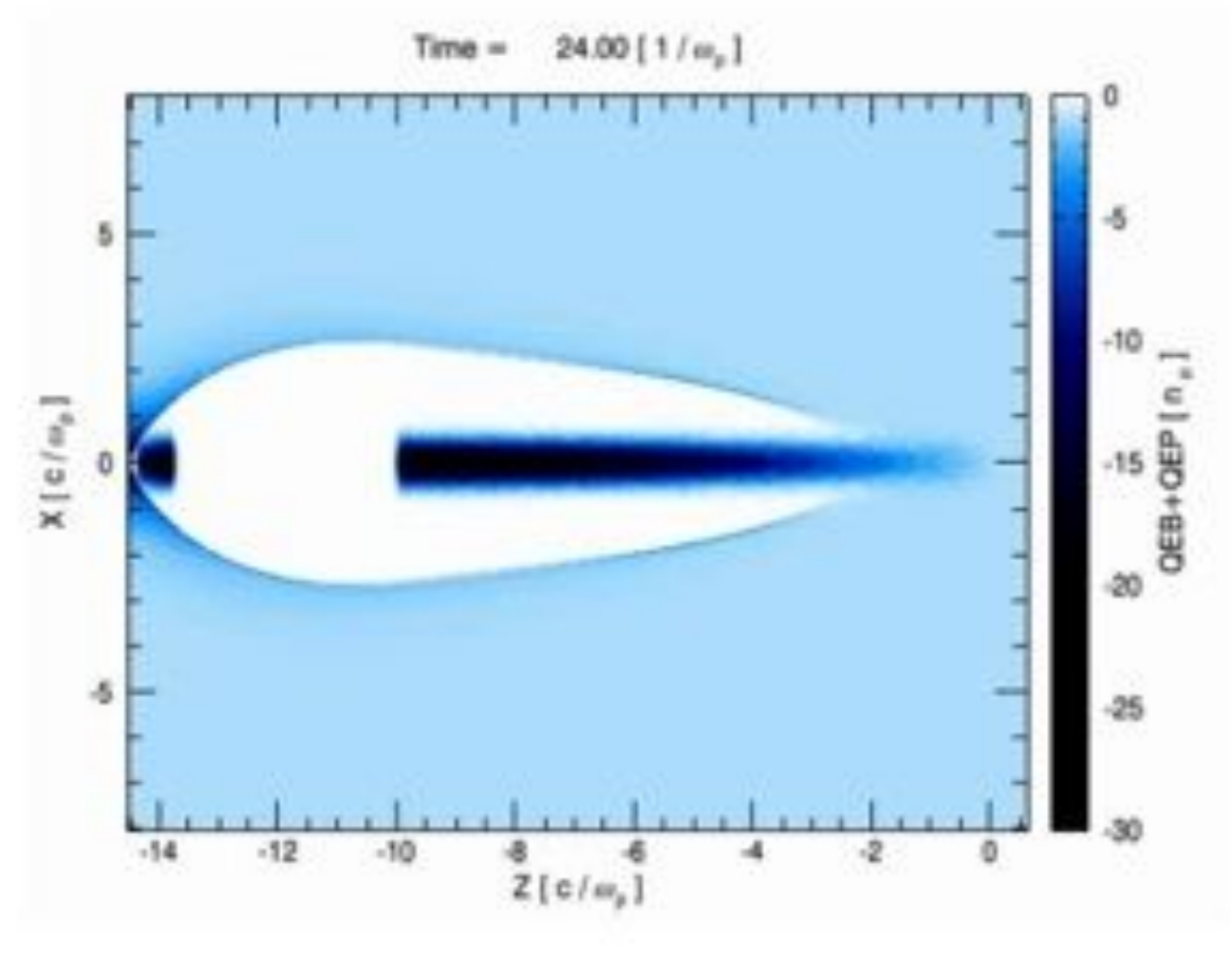}
\caption{OSIRIS simulation of a wedge shaped bunch (reproduced from \cite{Lu_PAC_2009}) demonstrating a transformer ratio of~5.7.}
\label{lupac}
\end{center}
\end{figure}

Hence, the energy loss rate of a \emph{single electron} in the drive beam (assuming it is ultrarelativistic so $v_z\approx c$) can be expressed as
$$
\frac{d}{dt} \gamma_b mc^2 \approx -N_b eE_{z1}(0)c\;.
$$
Likewise, the energy loss rate for a single electron in the witness bunch is
$$
\frac{d}{dt} \gamma_w mc^2 \approx -N_b eE_{z1}(w)c-N_w eE_{z1}(0)c\;.
$$
In this expression, the first term on the right hand side represents the electric field of the wake generated by the driver and the second term represents the wake of the witness bunch. If we sum these two expressions together weighted by the number of electrons in each bunch the total is the rate of change of total energy,
$$
N_b\frac{d}{dt} \gamma_b mc^2 + N_w\frac{d}{dt} \gamma_w mc^2 = \frac{d}{dt} ({\rm Total\; energy})\;,
$$
and therefore since clearly the total energy can't increase,
$$
N_b\frac{d}{dt} \gamma_b mc^2 + N_w\frac{d}{dt} \gamma_w mc^2 \leq 0\;,
$$
and so 
$$
N_b^2 eE_{z1}(0)c+N_bN_w eE_{z1}(w)c+N_w^2 eE_{z1}(0)c\geq0\;.
$$
Clearly, for there to be any energy gain by the witness bunch at all, the magnitude of $E_{z1}(w)$ must be negative, and so
$$
(N_b^2 +N_w^2) E_{z1}(0)\geq N_bN_w |E_{z1}(w)|\;.
$$
For a symmetric situation where the two bunches are the same charge, this leads to
$$
|E_{z1}(w)|\leq 2 E_{z1}(0)\;,
$$
or in other words the accelerating gradient at the bunch can be no more than double that of the drive bunch. Thus, the total accelerating gradient experienced by the witness bunch, $E_z(w)$ is
$$
E_z(w) \leq (2N_b - N_w)E_{z1}(0)\;.
$$
For an electron beam driver, the \emph{depletion length} can be formulated by setting the initial beam energy per particle $\gamma_b mc^2$ equal to the work done on an electron in the beam, i.e. 
$$
L_{pd} = \frac{\gamma_b mc^2} {eE_{z1}(0)N_b}\;.
$$
Since the maximum work done on an electron in the \emph{witness} beam is 
$$
\Delta\gamma_w mc^2 = eE_z(w)L_{pd}\;,
$$
we can combine these expressions to obtain
$$
\Delta\gamma_w mc^2  \leq \frac{(2N_b - N_w)}{N_b}\gamma_b mc^2\;.
$$
The \emph{transformer ratio}, $R_T$, is the ratio of the energy gain by the witness beam relative to that of the~driver beam, i.e.
$$
R_T = \frac{\Delta\gamma_w}{\gamma_b} = 2-\frac{N_w}{N_b}\;.
$$
It therefore quantifies the maximum energy transfer from driver to witness.
It can also be seen that it is 
$$
R_T = \frac{E_z(w)}{E_z(b)}\;,
$$
where $E_z(b)$ is the electric field at the drive beam. This upper limit can be overcome by asymmetric and shaped bunches. For a wedge shaped bunch of length $L$ \cite{Lu_PAC_2009},
$$
R_T\sim \frac{L}{\Lambda_0}\;.
$$
Figure \ref{lupac} shows a wedge shaped bunch from an OSIRIS simulation (reproduced from \cite{Lu_PAC_2009}) demonstrating a transformer ratio of 5.7.
The overall energy transfer efficiency $\eta$ is the transformer ratio weighted by the bunch charges, i.e. 
$$
\eta = \frac{N_w\Delta\gamma_w}{N_b\gamma_b} \leq \frac{N_w}{N_b}\left(2-\frac{N_w}{N_b}\right)\;.
$$
\subsection{Beamloading and energy spread considerations}
We have ignored the effect of the fields of the witness bunch so far. This is fine if the witness bunch charge is small, but if the witness bunch charge-per-unit-length approaches that of the plasma density perturbation, its effect on the wakefield will be significant. It is possible to flatten the field completely with an appropriately shaped bunch. This is known as \emph{beamloading}. Beamloading will result in the~electron beam being accelerated with equal field strength for all electrons even if the bunch is relatively long and therefore will preserve the beam energy spread.

\begin{figure}[htbp]
\begin{center}
\includegraphics[width = 0.3\textwidth]{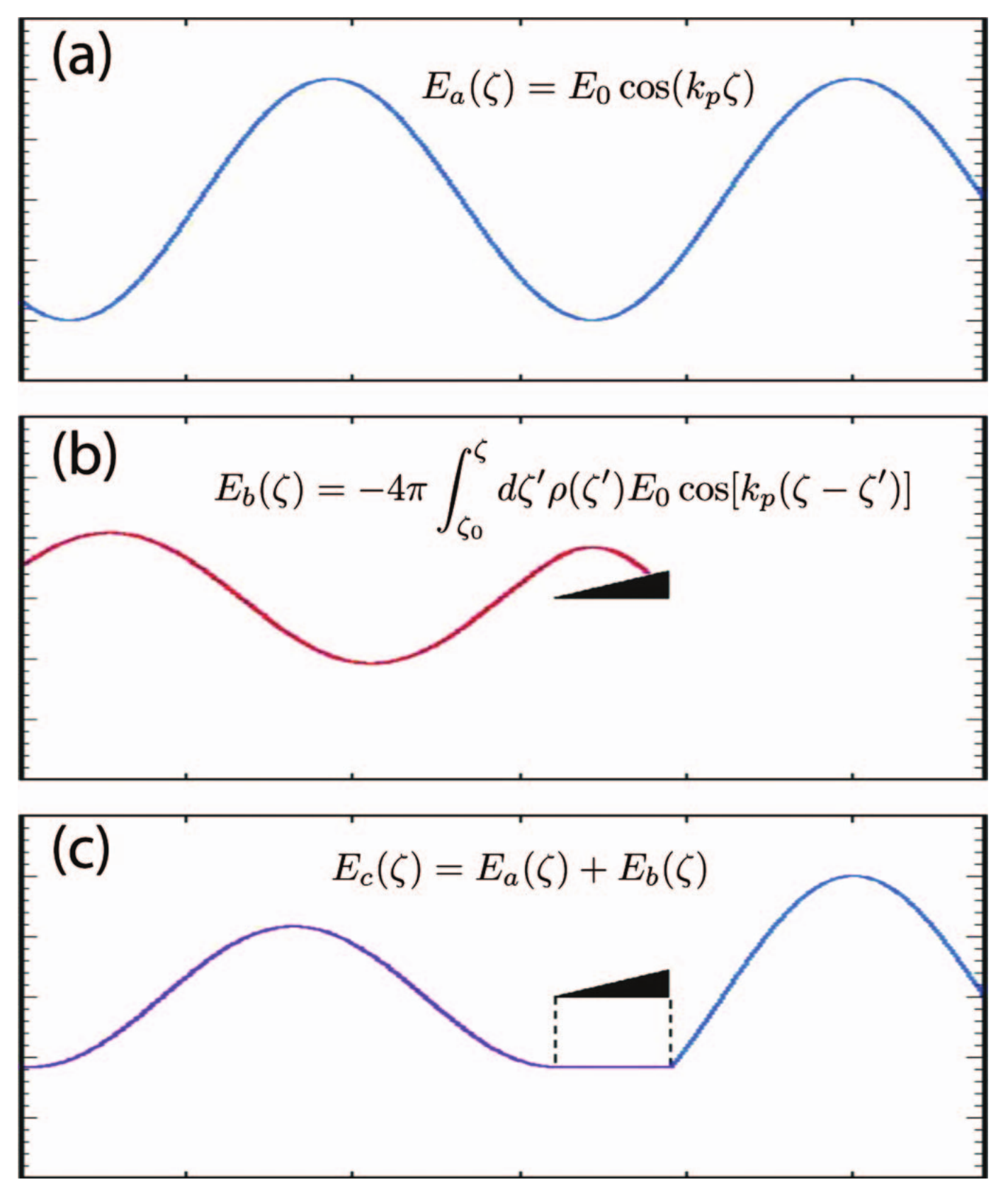}
\caption{Contributions of the wake due the driver and that of an appropriately shaped witness bunch can lead to a~flattened electric field. (Reproduced from ref. \cite{Tzoufras_2009}.)}
\label{tsoufras}
\end{center}
\end{figure}

 For ideal flattening of the field, i.e., the electrons in the bunch experience an identical field throughout, the bunch charge density per unit length must equal the background ion charge density per unit length. In three-dimensions, Fig. \ref{tsoufras}, taken from Ref. \cite{Tzoufras_2009}, shows how the contributions of the wake due the driver and that of an appropriately shaped witness bunch can lead to a flattened electric field in the~region of the bunch. 
\section{Summary}
Having an accelerating field structure move at $v_p$ results in upshift of both the maximum energy gain, compared to a stationary accelerating structure with identical potential, and accelerator length by dephasing, by a factor $2\gamma_p^2$, in the most general sense. 

For  laser driven plasma wakefield accelerators, since the dispersion in plasma results in a Lorentz factor $\gamma_p\sim \omega_0/\omega_p$, the accelerator length scales as the plasma number density to the three halves power, $\propto n_0^{3/2}$. The maximum energy scales proportionally to the plasma number density, $n_0$. This means that the accelerating gradient scales as $1/\sqrt{n_0}$, which implies that for the highest energy gains for high energy physics, staging will be required.

For the beam driven case, the maximum energy gain is limited by driver energy loss under realistic conditions. 
The acceleration length scales as the beam energy and inversely to the decelerating field strength,
$$
L_{pd} = \frac{\gamma_b mc^2} {e|E_{z}(b)|}\;,
$$
and the maximum energy gain is limited by the transformer ratio,
$$
R_T = \frac{E_z(w)}{E_z(b)}\;.
$$
The energy spread of the accelerated beam is affected by phase space rotation, the fields of the bunch (with beam loading required for flattening the field). Low energy spreads can be achieved for correct extraction timing / accelerated charge profile / localisation of trapping in wake phase.

\end{document}